\documentclass{aa}  
\usepackage[varg]{txfonts}
\usepackage{natbib}
\bibpunct{(}{)}{;}{a}{}{,}
\usepackage{graphicx}
\usepackage[breaklinks=true,colorlinks=true,linkcolor=black,anchorcolor=black,citecolor=blue,filecolor=black,menucolor=black,runcolor=black,urlcolor=black]{hyperref}
\usepackage{graphicx}
\usepackage{multirow,bigdelim}
\usepackage{caption}
\usepackage{float}
\usepackage{url}
\usepackage{arydshln}
\usepackage{multirow}
\usepackage{subfig}
\begin{document}

\title{Differences in the solar cycle variability of simple and complex active regions during 1996 - 2018}
  
  \author{S. Nikbakhsh\inst{1}
  \and E. I. Tanskanen\inst{1} 
  \and M. J. Käpylä\inst{2,3}
  \and T. Hackman \inst{4}} 

\institute{ReSoLVE Centre of Excellence, Department of Electronics and Nanoengineering, Aalto University, 02150 Espoo, Finland  \\
\email{shabnam.nikbakhsh@aalto.fi}
  \and Max Planck Institute for Solar System Research, Justus-von-Liebig-Weg 3, 37077 Göttingen, Germany
    \and ReSoLVE Center of Excellence, Department of Computer Science, Aalto University, 02150 Espoo, Finland
    \and Department of Physics, P.O. Box 64, FI-00014 University of Helsinki, Finland}

\abstract
{}{Our aim is to examine the solar cycle variability of magnetically simple and complex active region.}{We studied  simple ($\alpha$ and $\beta$) and complex ($\beta\gamma$ and $\beta\gamma\delta$) active regions based on the Mount Wilson magnetic classification by applying our newly developed daily approach. We analyzed the daily number of the simple active regions (SARs) and compared that to the abundance of the complex active regions (CARs) over the entire solar cycle 23 and cycle 24 until December 2018.}{We show that CARs evolve differently over the solar cycle from SARs. The time evolution of SARs and CARs on different hemispheres also shows differences, even though on average their latitudinal distributions are shown to be similar. The time evolution of SARs closely follows that of the sunspot number, and their maximum abundance was observed to occur during the early maximum phase, while that of the CARs was seen roughly two years later. We furthermore found that the peak of CARs was reached before the latitudinal width of the activity band starts to decease.}{Our results suggest that the active region formation process is a competition between the large-scale dynamo (LSD) and the small-scale dynamo (SSD) near the surface, the former varying cyclically and the latter being independent of the solar cycle. During solar maximum, LSD is dominant, giving a preference to SARs, while during the declining phase the relative role of SSD  increases. Therefore, a preference for CARs is seen due to the influence of the SSD on the emerging flux.}

\maketitle 

\section{Introduction}
\label{sec:intro}

Solar active regions (ARs) appear on the Sun's photosphere in a variety of sizes, shapes, and magnetic complexities or magnetic topologies \citep{Howard1989}. A considerable number of studies have used the magnetic complexity of ARs as a measure of their activity \citep{Cortie1901, Waldmeier1938, McIntosh1990, Moon2016}. In 1919,  \citet{Hale1919} introduced the Mount Wilson (or Hale) Classification, which groups ARs according to the distribution of magnetic field topology. Three major complexity classes were included in this scheme: unipolar ($\alpha$), bipolar ($\beta$), and multipolar ($\gamma$). Since 1919, the original classification has been modified several times, but the most profound change occurred when \citet{Kunzel1960} introduced a new complexity class $\delta$ in 1960. The current version of the Mount Wilson Classification groups ARs into the five major classes (or configurations): $\alpha$, $\beta$, $\beta\gamma$, $\gamma,$ and $\delta$ \citep{Lidia2015}. Table \ref{table:mount} presents the categorizing rules for grouping the magnetic complexity of ARs  according to the Mount Wilson Classification. This scheme has been an effective and widely-used means of studying ARs \citep{Tang1984, Ireland2008, Stenning2013} because of its relatively simple approach to classification.

\begin{table}
\caption{Mount Wilson Classification rules for grouping solar ARs according to their magnetic complexity.}
\label{table:mount}
\centering
\begin{tabular}{| c | l |}
\hline
Class symbol &  Classification rules \\
\hline
$\alpha$ & A unipolar sunspot group \\ 
             & \\ 
$\beta$  & Bipolar sunspot group with\\
              & a distinct section between polarities  \\ 
              & \\
$\beta\gamma$ & A bipolar sunspot group \\
             &  whose opposite polarities cannot  \\
             & be separated with a continuous line  \\
             & \\
$\gamma$ & A complex sunspot group with\\
            &  mixed positive and negative polarities,\\
            &  which cannot be identified as a bipolar  \\
           & \\
$\delta$ & A complex sunspot group \\
           &  that contains opposite polarity  \\
           &  umbrae within the same penumbra \\                  
\hline
\end{tabular}
\tablefoot{Reprinted from {\citet[pp. 80]{Lidia2015}.}}
\end{table}

 It is to be noted that each active region may be classified according to its magnetic complexity with one or a combination of two major classes, for instance $\beta$ or $\beta\gamma\delta$.  In addition, the magnetic complexity of an AR  is not necessarily constant, as it might change from one class to another during a region's lifetime \citep{Tang1983}. ARs frequently host various types of eruptive phenomena, including solar flares and coronal mass ejections (CMEs), and the flare and CME production rate differs between ARs \citep{Zhongxian1993, Sammis2000, Takizawa2015}.

 It is commonly acknowledged that flares and CMEs are among the major drivers of space weather although they do not always occur within ARs \citep{Gonzalez2016, Mursula2013,Subramanian2001}. 
Nevertheless, predicting the flare and CME production of ARs can profoundly contribute to space weather forecasting \citep{Wang2015}. To date, various studies have investigated the characteristics of ARs to estimate their flare and CME productivity \citep{Smith1968}.  For instance, \citet{Chen2011} conducted a statistical study of the role of magnetic complexity in the CME production rate of ARs. They found that more complex ARs, for example the $\beta\delta$ structure, are more CME productive than simple ARs (e.g., $\alpha$). \citet{Guo2014} also reported that  most of the recorded X-class flares in their sample originated in ARs in $\beta\gamma\delta$ class, although they showed that the complexity of ARs had no significant effect on the kinematic properties of CMEs. 

One of the most comprehensive early studies using the Mount Wilson Classification was \citet{Hale1938}, which investigated the magnetic complexity of 2174 ARs observed by the Mount Wilson telescopes during the period 1915 to 1924 covering the majority of solar cycle (SC) 15 and the beginning of SC 16. Since the complexity class of ARs might change each day, they used the average of the complexity class for each AR appearing in the sample. Among all the ARs that they studied, the majority had a simple complexity and only $4 \%$ belonged to a more complex class (the $\delta$ magnetic configuration was introduced later, in 1960). Moreover, their results suggest no significant correlation between the distribution of AR complexity classes and the sunspot cycle.

In a more recent study, \citet{Amiee2016} reported a similar statistical analysis of 5468 ARs for the period of 1992 to 2015. This time interval covers the end of SC 22, the entirety of SC 23, and the majority of SC 24. They analyzed the NOAA Active Region (NAR) database, which is provided by the National Oceanic and Atmospheric Administration (NOAA). The researchers selected the complexity class of each AR present in the NAR database from the day each region reached its maximum sunspot area during its lifetime. They also demonstrated that the majority of ARs appeared with a simple complexity class ($\alpha$ and $\beta$) , whereas $16.24\%$ had a complex configuration. Furthermore, in contrast to the previous study, they found the magnetic complexity of ARs varied with the solar cycle.

Many other studies have investigated the magnetic complexity of ARs using modified versions of the Mount Wilson Classification, but to our knowledge no study has investigated the daily abundance of magnetic complexities and its variation over time. In this paper we demonstrate that by using the daily complexity number of ARs, important information about the lifetime and magnetic complexity evolution of ARs can be achieved. Furthermore, this approach enables us to compare our results to the solar magnetic activity and its evolution.

\section{Data and method}
\label{sec:DATA}

In order to study the magnetic complexity of ARs, we analyzed the Solar Active Region Summary (SRS) list available on the Heliophysics Integrated Observatory (HELIO) website.\footnote{\url{http://www.helio-vo.eu/}}  The SRS list reports the comprehensive daily details of the ARs present on the visible solar disk for the preceding day, such as total area, heliographic longitude and latitude, Mount Wilson classification, and the number of sunspots in each region. These data are obtained from the daily sunspot report of the Solar Optical Observing Network (SOON) under a joint project between NOAA and the US Air Force (USAF), 2nd Weather Squadron \citep{Balamjan2016}. The SOON telescope images the Sun at the H$\alpha$ wavelength to locate sunspots on the photosphere. The telescope is also supplied with a longitudinal magnetograph to map the Sun's magnetic field by utilizing the Zeeman effect  \citep{Zeeman1897, Spencer1965}.

\begin{figure}[h]
\resizebox{\hsize}{!}{\includegraphics{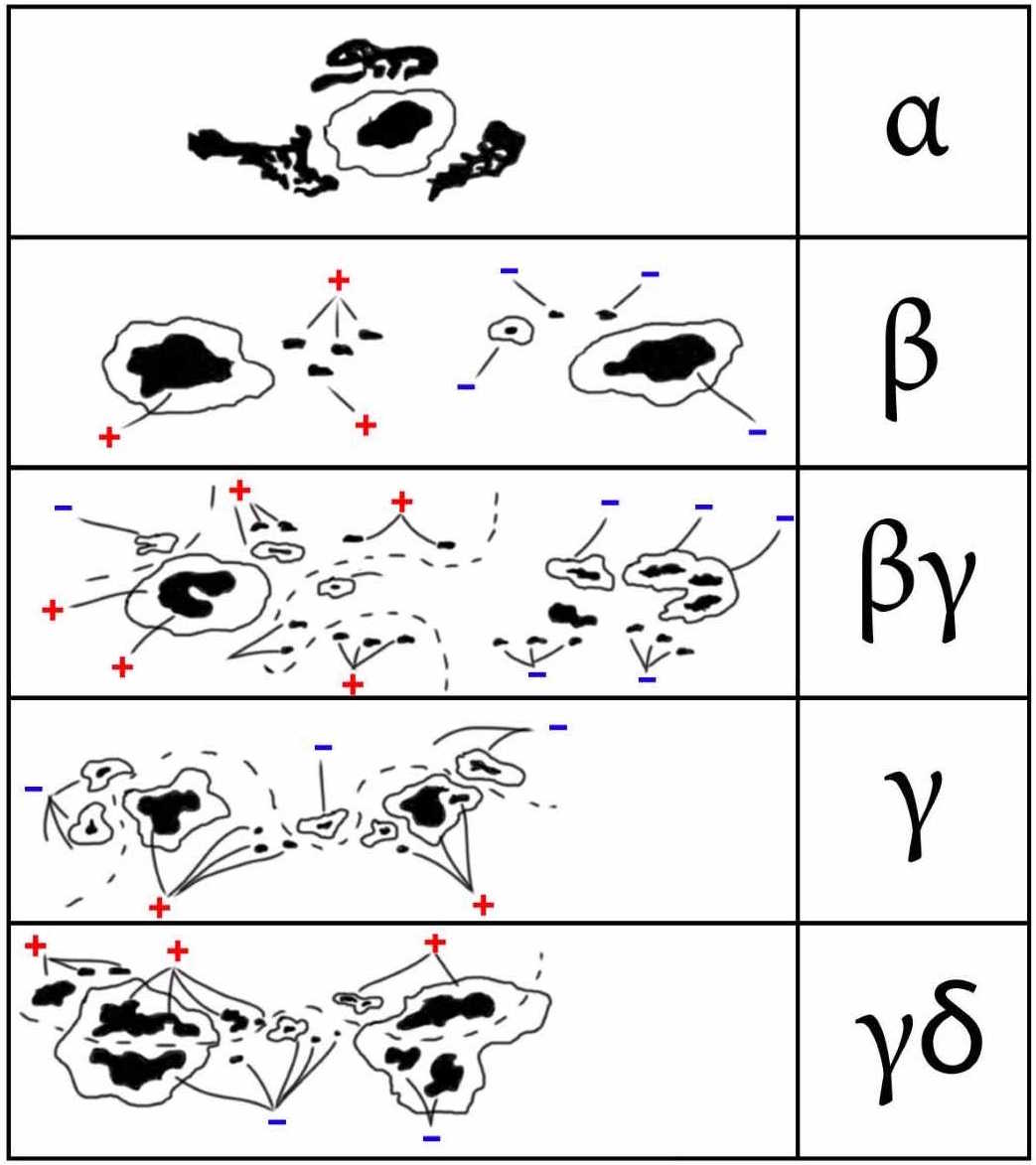}}
\caption{Examples of the SOON’s classification system for the identification of magnetic complexity of ARs  based on  the Mount Wilson Classification. The image is taken from \citet{Andrus2013}}
\label{fig:SOON}
\end{figure} 

The magnetograph analyzes the Fe I spectral line at 8468  \AA\  in order to evaluate the magnetic polarity of the AR (positive or negative) as well as the line-of-sight component of the magnetic field strength. As a result, SOON’s magnetograph produces line-of-sight magnetic field maps of the Sun that are called magnetograms, with a resolution of five-arcseconds pixels \citep{ISOON}. Then the Mount Wilson Classification is applied to determine the magnetic configuration of the ARs present in the magnetograms. Among all the complexity classes available in the Mount Wilson Classification, SOON only reports the following classes: $\alpha$,$\,$ $\beta$,$\,$ $\beta\delta$,$\,$ $\beta\gamma$,$\,$ $\beta\gamma\delta$,$\,$ $\gamma,$ and $\gamma\delta$ \citep{Andrus2013}. Figure \ref{fig:SOON} provides several examples of magnetic complexity types with the identification system applied by SOON. The illustration shows that more complex ARs, for example $\beta\gamma$, are classified with multiple major classes. The plus and minus signs show the positive and negative polarities of ARs, respectively.
 
We analyzed the daily magnetic complexity and latitude data of ARs available in the SRS list, for the period January 1996 to December 2018. This time interval includes the whole of SC 23 (Jan. 1996 \textendash\,Dec. 2008) and the majority of SC 24 (Jan. 2009 \textendash\,Dec. 2018). The SRS list assigns one complexity class to each individual AR per day. As explained in the introduction, the magnetic complexity of an AR might change from one class to another during its lifetime. We analyzed the data by applying the daily magnetic complexity approach, which includes all associated complexity classes in each AR. For example, in the case of an AR that appeared on the visible disk for five days with $\alpha$ configuration, we added five $\alpha$ to the final calculation of the abundance of various magnetic complexity classes. Alternatively, if an AR emerged with $\alpha$ configuration for two days and then transformed into $\beta$ configuration for three days, we would add two $\alpha$ and three $\beta$ to the final calculation. In this way, we combined the information of both the lifetime and magnetic complexity of ARs in our analyses. We mostly focused on studying the most common complexity classes among all classes that are identified in the SRS list: $\alpha$ , $\beta$ , $\beta\gamma,$  and $\beta\gamma\delta$. Furthermore, we divided these classes into two distinct groups: simple ARs (SARs), including $\alpha$  and $\beta$  classes, and complex ARs (CARS), including $\beta\gamma$  and $\beta\gamma\delta$. We also compared our results with the cyclic variation of the sunspot number. 

\section{Results}
\label{sec:RESULTS} 

\begin{figure}[h]
\resizebox{\hsize}{!}{\includegraphics{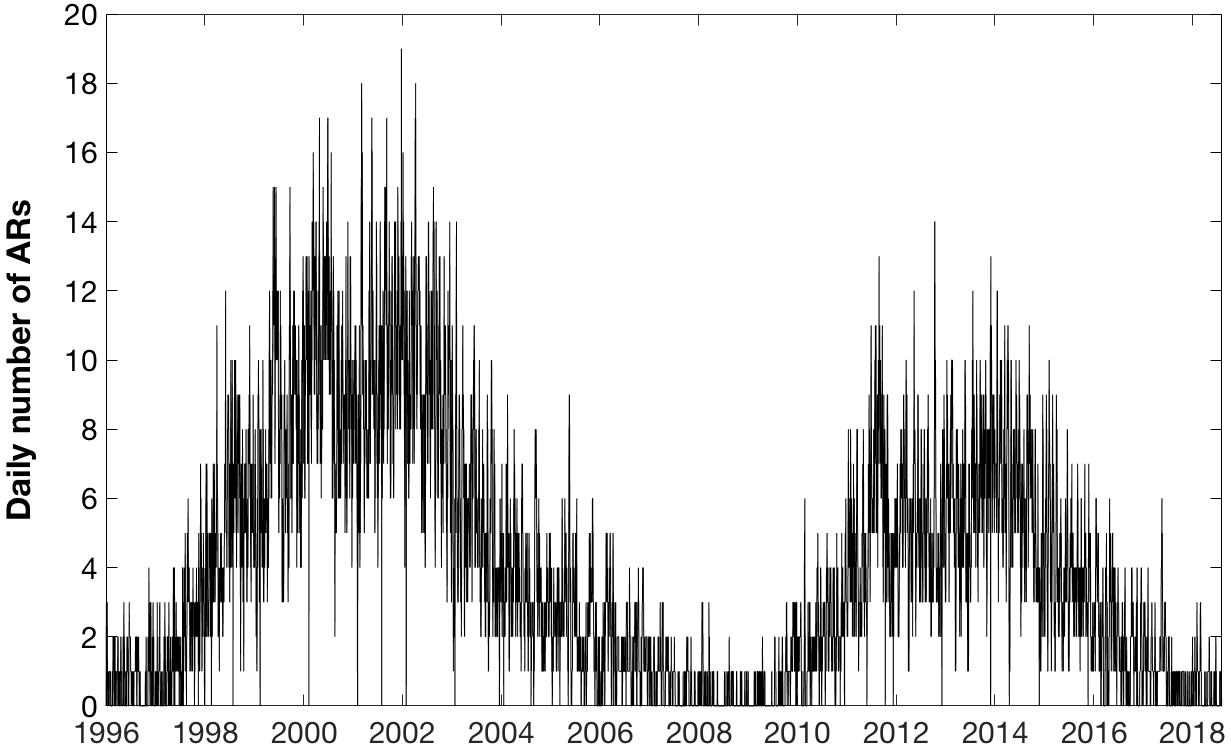}}
\caption{Daily number of ARs from January 1996 to December 2018.}
\label{fig:Daily}
\end{figure}

We studied 4797 distinct ARs from the SRS list for the period January 1996 to December 2018. The total  daily count of ARs, including all complexity classes ($\alpha$, $\beta$, $\beta\gamma$, $\beta\gamma\delta$, $\beta\delta$,$\gamma,$ and $\gamma\delta$) using our daily magnetic complexity approach was 33496. Figure \ref{fig:Daily} presents the variation in the daily number of ARs obtained according to this approach. Our approach is different from that of \citet{Amiee2016} in the sense that it results in several complexity data points per AR as they evolve, while \citet{Amiee2016} determine only one complexity data point per AR, taken from the time when the area of the AR reached its maximum. In practice, this means that our statistic will emphasize long‐lived ARs more than short‐lived ones.

Next, we calculated the abundance of each complexity class in the sample by applying our approach from 1996 to 2018. Table \ref{tab:result} presents the total count and relative abundance of each complexity class in the sample. The data in Table \ref{tab:result} show that the majority of ARs are in $\beta$ class ($57.57 \%$) and the next most common class is $\alpha$, which accounts for $30.73\%$ of ARs. The $\beta\gamma$ and $\beta\gamma\delta$ configurations, which are among the complex ARs, in turn account for $11.68 \%$. The rest of the configurations (i.e., $\beta\delta$, $\gamma$, and $\gamma\delta$) only occurred with a relative abundance of $0.51 \%$ and we excluded them entirely from further analysis in this paper.

\begin{table}[h]
\caption{Total and relative abundances of each complexity class according to the daily number of ARs from January 1996 to December 2018.}
\label{tab:result}
\centering
\begin{tabular}{| c | c c r c c |}
\hline  
Complexity & Count & Relative abundance & & &\\
 & [number] & [$\%$]  & & &\\
\hline       
$\boldsymbol{\alpha}$ & 10296 & 30.73 & \rdelim\}{2}{2.5mm}[\textbf{88.30}] & &\\
$\boldsymbol{\beta}$  & 19284 & 57.57 &   &  &\\ 
\hdashline
$\boldsymbol{\beta\gamma}$ & 2919  & 8.71 & \rdelim\}{2}{2.5mm}[\textbf{11.68} ]& &\\
$\boldsymbol{\beta\gamma\delta}$& 997  & 2.97 & & &\\   
\hline
 $\beta\delta$  & 166 & 0.49& & &\\     
$\gamma$& 4  & 0.01& & &\\  
$\gamma\delta$& 5  & 0.01& & &\\  
\hline
Total count & 33671 & & & &\\
\hline                                   
\end{tabular}
\end{table} 

\begin{figure}[h]
\resizebox{\hsize}{!}{\includegraphics[width=1\linewidth]{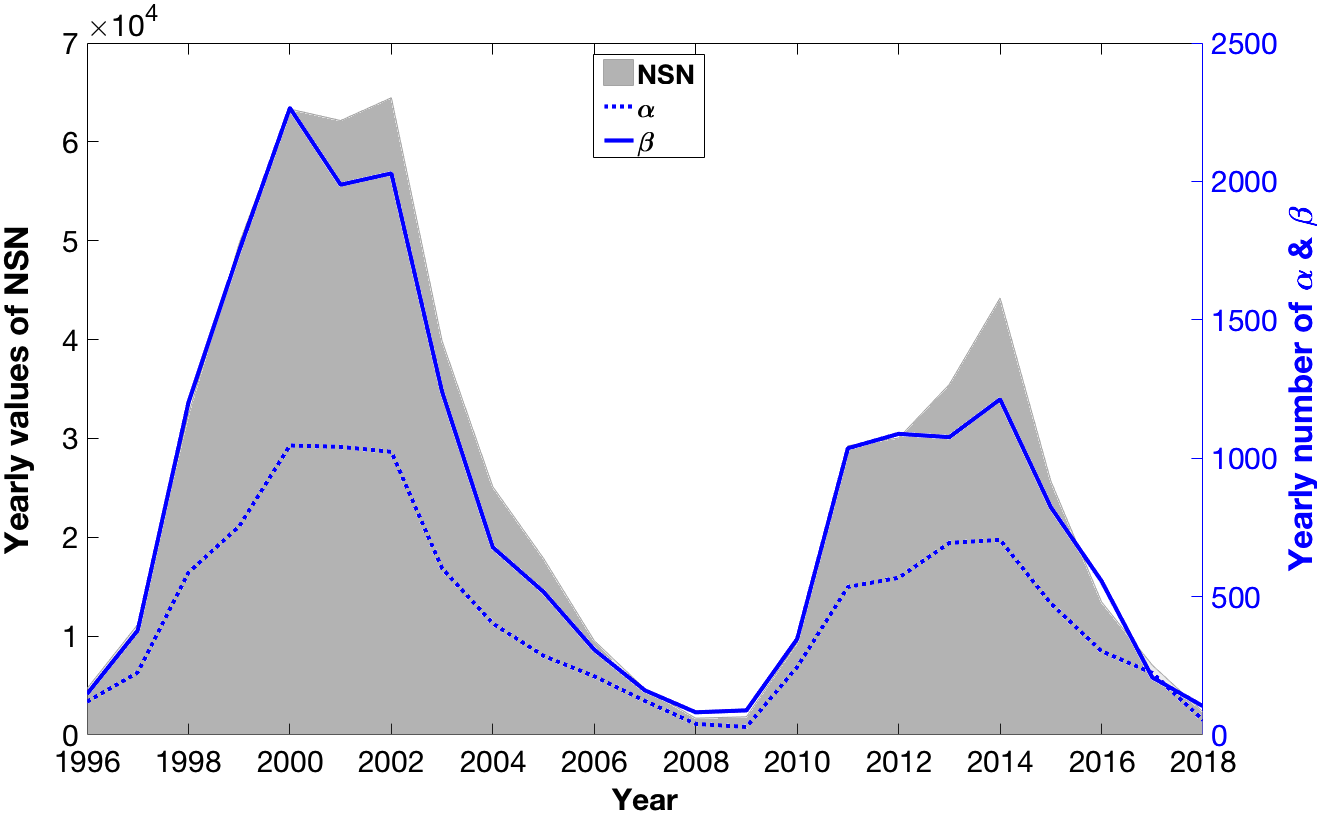}}
\resizebox{\hsize}{!}{\includegraphics[width=1\linewidth]{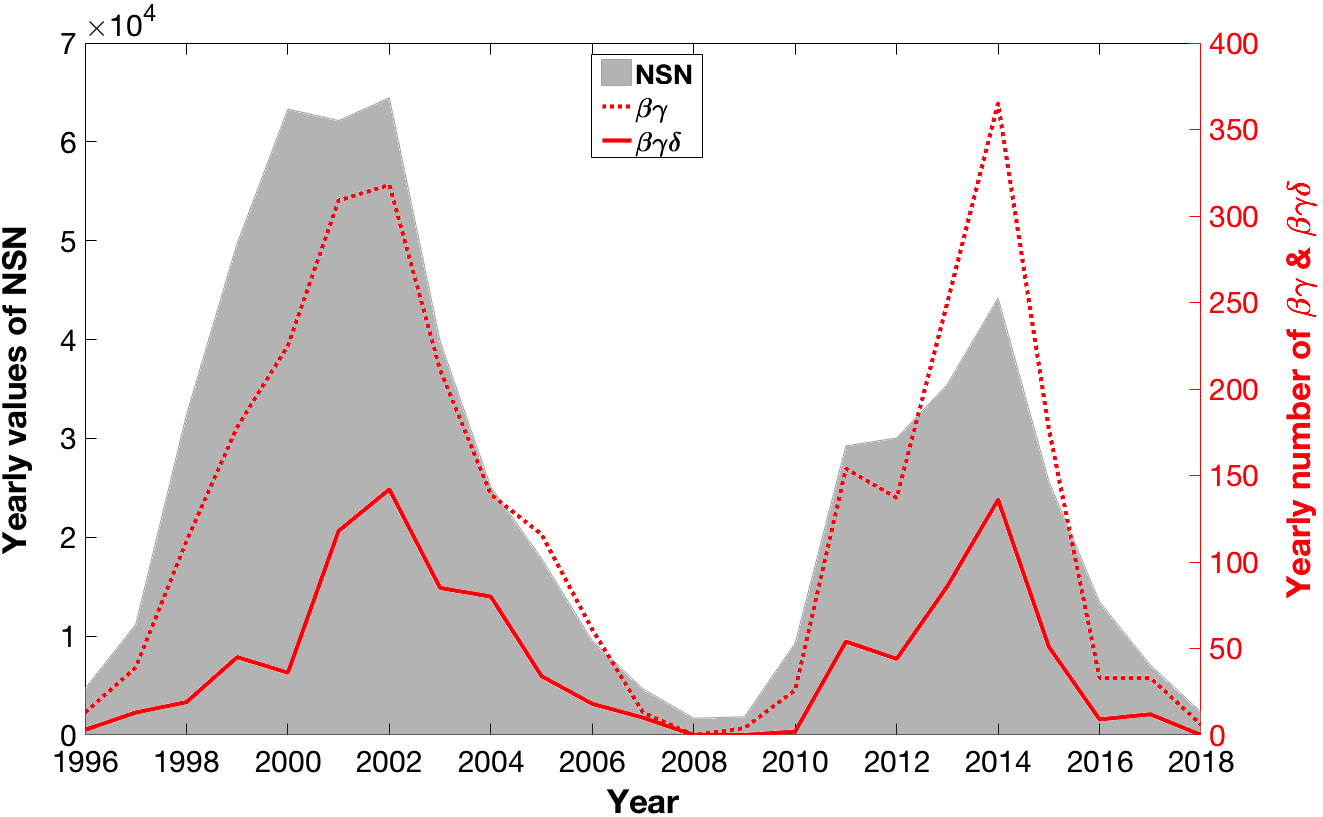}}
\caption{Yearly numbers of the most common magnetic complexities in comparison with the yearly average values of the NSN (gray shaded area). Upper panel: Yearly number of the $\alpha$ and $\beta$ classes are shown with the blue solid and dotted lines, respectively. Lower panel: Yearly values of the $\beta\gamma$ (red dotted line) and $\beta\gamma\delta$ classes (red solid line)}
\label{fig:Yearly} 
\end{figure}

Figure \ref{fig:Yearly} shows the yearly numbers of $\alpha$, $\beta$, $\beta\gamma,$ and $\beta\gamma\delta$ class ARs. To compare our results with the sunspot number, we chose the Boulder/NOAA sunspot number \citep{Hathaway2015}, which we call NSN, for the period of 1996 to 2018. It is worth mentioning that the NSN closely follows the international sunspot number (version 2.0, \citet{Brussel2016}). 

The upper panel Figure \ref{fig:Yearly} shows that the yearly number of $\beta$ ARs closely follows the same trend as the yearly values of the NSN during both SCs 23 and 24. In addition, this class shows a double peak behavior during the maximum phase in both cycles. However, there is a dramatic decrease in the yearly number of $\beta$ ARs during the maximum phase from SC 23 to 24. Furthermore, the yearly number of $\alpha$ ARs is relatively constant during the maximum phase in both cycles and does not show the same trend as the NSN during these phases. 

The yearly numbers of $\beta\gamma$ and $\beta\gamma\delta$ regions are presented in the lower panel of Fig. \ref{fig:Yearly}. The plot shows that the $\beta\gamma$ regions follow a double peak pattern only in SC 24, whereas the $\beta\gamma\delta$ regions show a double peak behavior in both cycles. Moreover, comparison of the two plots clearly indicates that the yearly numbers of $\beta\gamma$ and $\beta\gamma\delta$ regions are almost equal during the maximum phases in both cycles. 

\subsection{Monthly numbers of active regions}
\label{subsec:monthly}
\begin{figure}[h]
\resizebox{\hsize}{!}{\includegraphics{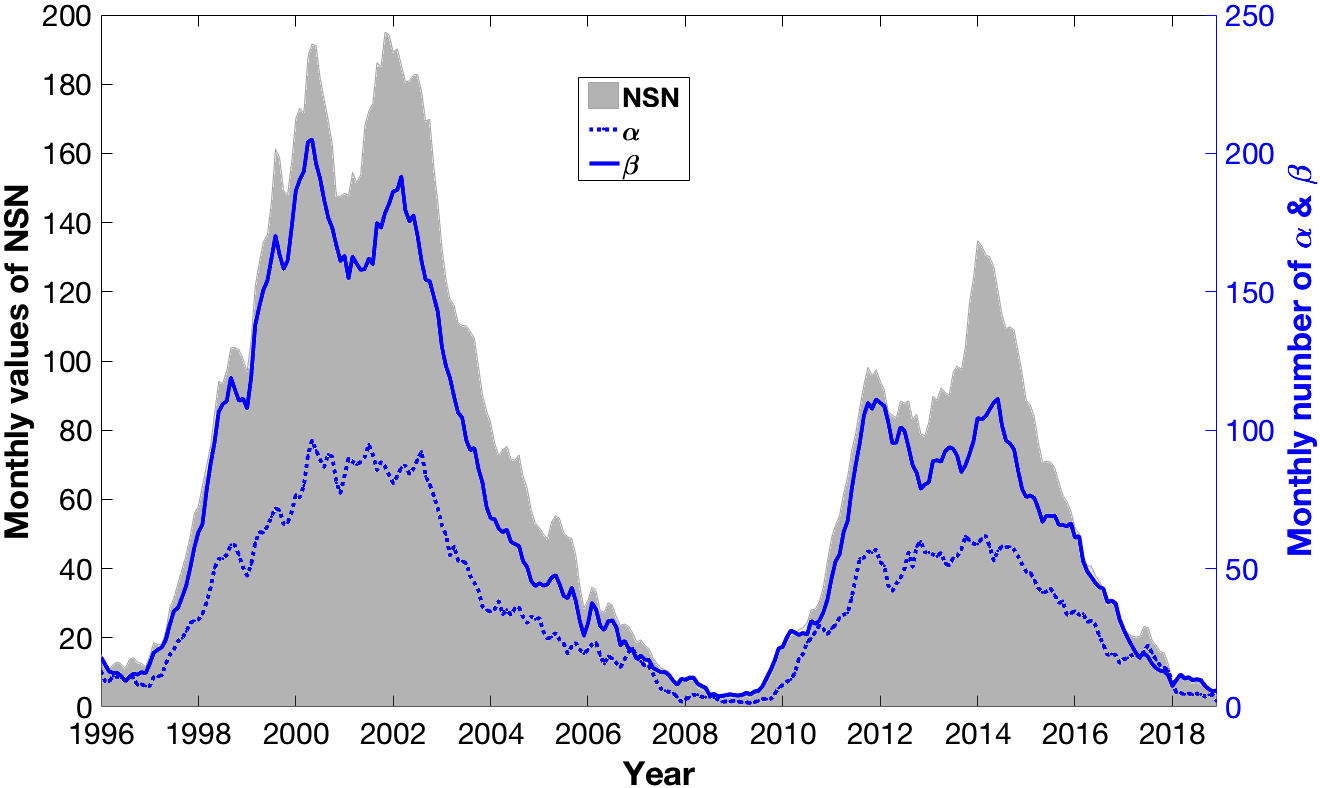}}
\resizebox{\hsize}{!}{\includegraphics{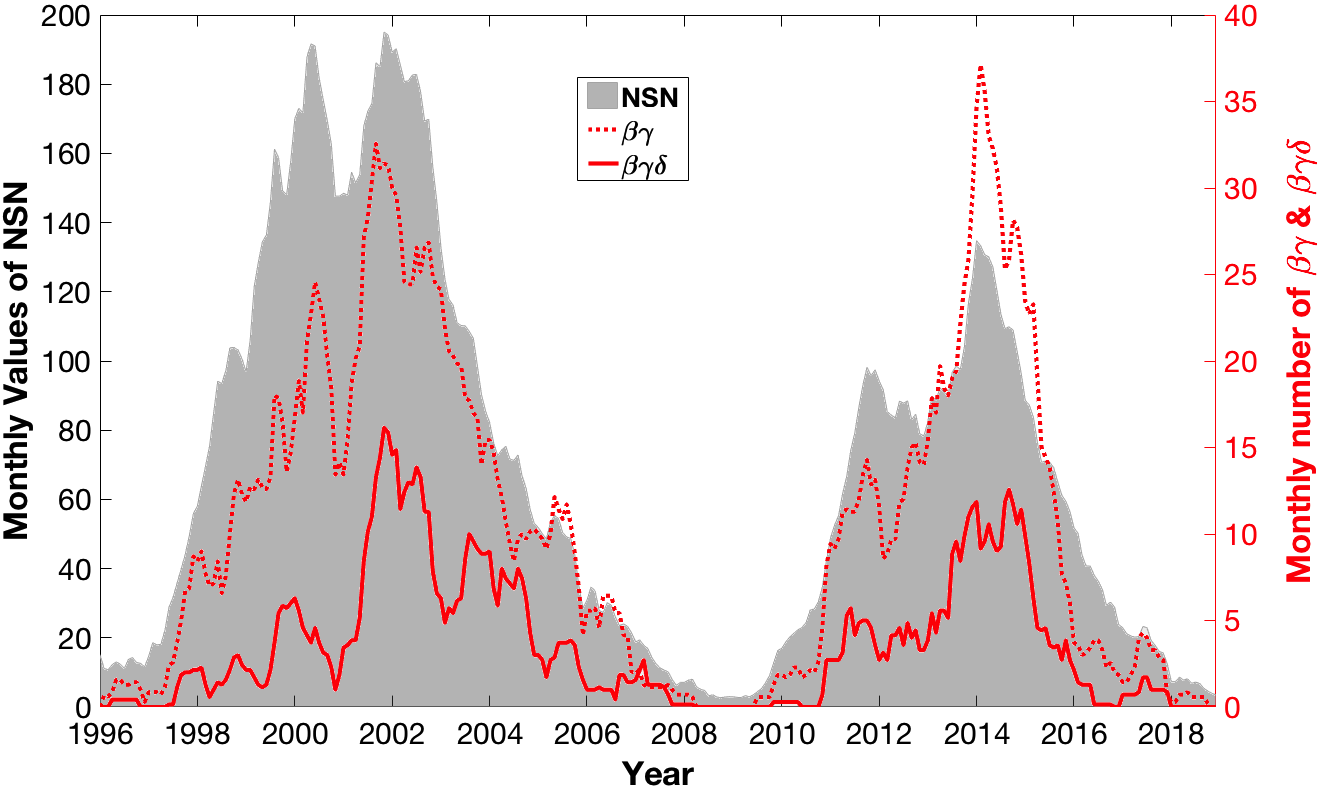}}
\caption{Monthly number of ARs in comparison with the monthly average values of the NSN (filled gray shaded areas in both panels). Upper panel: The monthly number of $\alpha$ and $\beta$ classes are displayed with the blue dotted and solid line, respectively. Lower panel: The monthly number of the $\beta\gamma$ (dotted line) and $\beta\gamma\delta$ (solid line) configuration.}
\label{fig:Monthly}
\end{figure}

The monthly number for each magnetic complexity was computed and the results were compared with the monthly average value of the NSN (Fig. \ref{fig:Monthly}). All data have been smoothed using a seven-month moving average.

The monthly number of $\alpha$ and $\beta$ ARs are shown in the upper panel of Fig. \ref{fig:Monthly}. This plot demonstrates that the monthly number of $\beta$ ARs closely follows the monthly values of the NSN and shows a double peak behavior during the maximum phase in both SCs 23 and 24; in each cycle, the two peaks are very similar in height, although the first peak is slightly higher. This trend was previously observed in the yearly number of $\beta$ ARs (see the upper panel of Fig. \ref{fig:Yearly}). Moreover, the monthly number of $\alpha$ ARs remains relatively constant during the maximum phase in both SCs 23 and 24. The same pattern can be seen in the yearly number of $\alpha$ ARs in the upper panel of Fig. \ref{fig:Yearly}. 

The monthly number of $\beta\gamma$ and $\beta\gamma\delta$ regions are presented in the lower panel of Fig \ref{fig:Monthly}. Both classes show a double peak behavior during the maximum phase in SCs 23 and 24; in each cycle, the two peaks are different in height and the second peak is higher. Furthermore, comparison of the two panels in Fig. \ref{fig:Monthly} clearly shows that the monthly number of $\beta$ regions reaches its maximum during the first NSN peak in SCs 23 and 24, whereas the monthly numbers of $\beta\gamma$ and $\beta\gamma\delta$ regions are highest later, during the second NSN peak in each cycle.

\subsection{Simple and complex active regions}
\label{subsec:simVscom}
We selected $\alpha$, $\beta$, $\beta\gamma,$ and $\beta\gamma\delta$ ARs from the SRS list and sub-categorized them into two groups: simple ARs (SARs) including $\alpha$ and $\beta$ and complex ARs (CARs) including $\beta\gamma$ and $\beta\gamma\delta$.  

\begin{figure}[h]
\resizebox{\hsize}{!}{\includegraphics{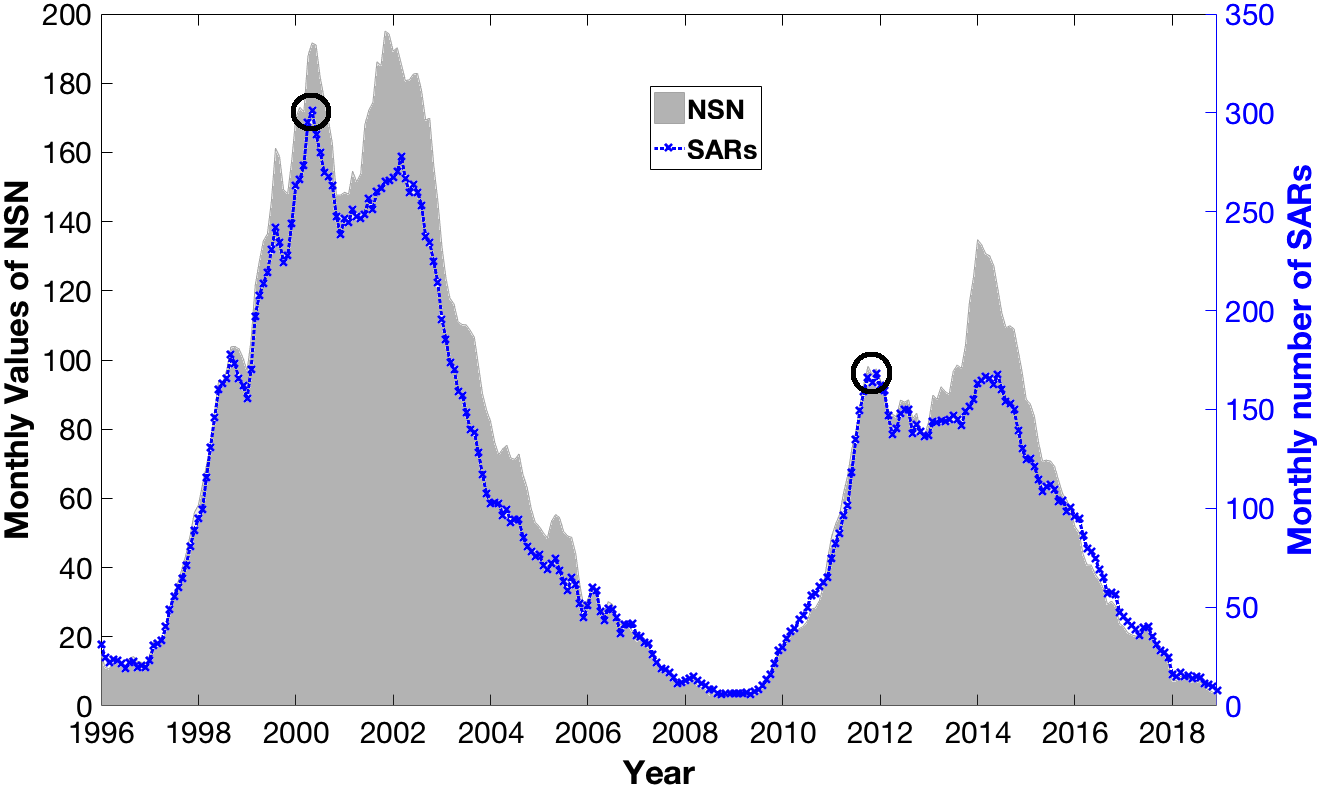}}
\resizebox{\hsize}{!}{\includegraphics{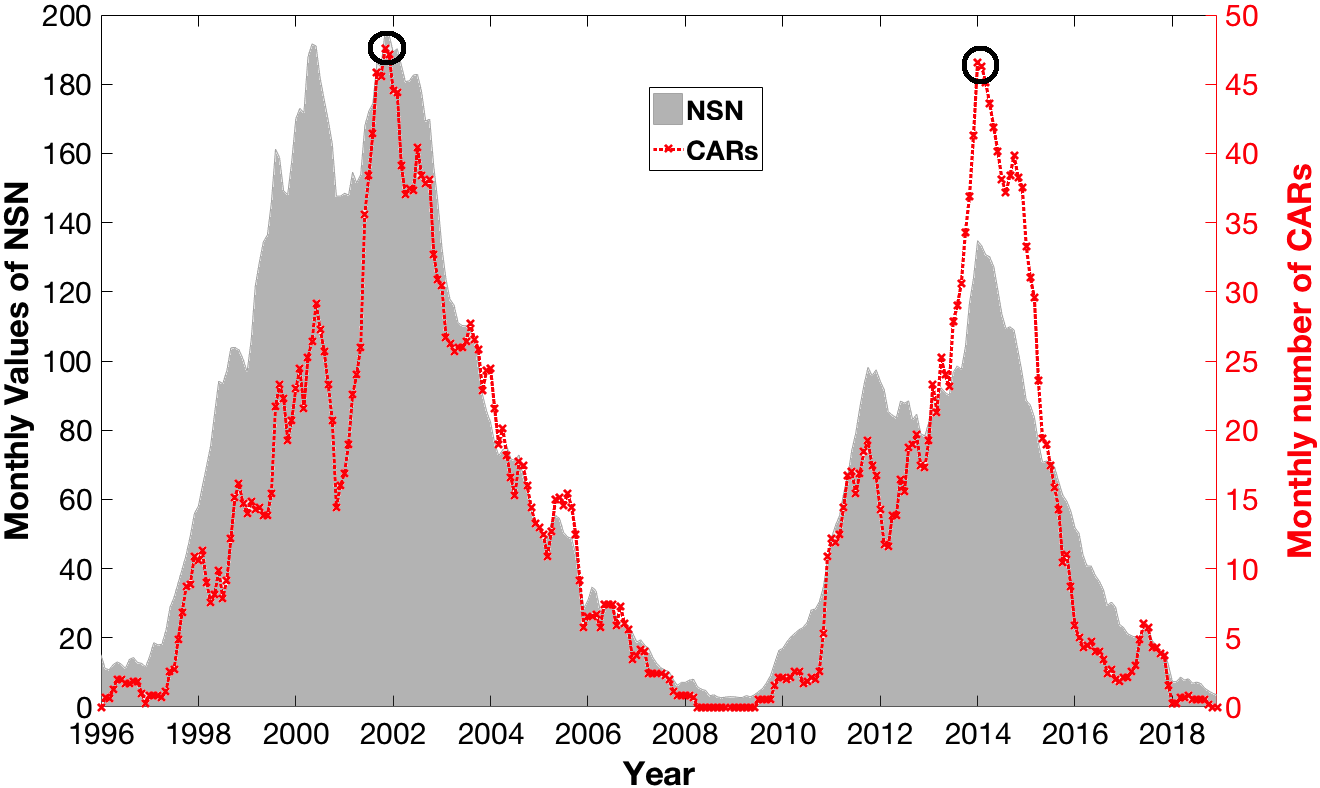}}
\caption{Monthly number of SARs (upper panel) and CARs (lower panel) in comparison with the monthly average values of the NSN (gray shaded areas). The maximum values of SARs and CARs are marked by circles in each cycle.}
\label{fig:SARs_CARs}
\end{figure} 

Figure \ref{fig:SARs_CARs} shows the monthly number of SARs (upper panel) and CARs (lower panel) against the monthly average values of the NSN (gray shaded areas in both panels) from January 1996 to December 2018 (The data have been smoothed by a seven-month running average). The monthly number of SARs in the upper panel displays a very similar trend as the monthly values of the NSN, including a double peak behavior during the maximum phase in both SCs 23 and 24. The Pearson correlation coefficient between SARs and the NSN values is $r = 0.99$. The lower panel of Fig. \ref{fig:SARs_CARs} presents the monthly number of CARs versus the monthly values of the NSN ($r = 0.87$). A double peak behavior is seen in the number of CARs during the maximum phase in both SCs 23 and 24, while the second peak is higher than the first one in each cycle.  

\begin{table}[h]
\caption{Monthly number of SARs and CARs for the period of two years before and after their maximum values in SCs 23 and 24.}
\label{tab:maxima}
\centering
\begin{tabular}{| c | c | c |  c |}
 \hline
 & Peak SARs & Peak SARs & Rate  \\  
 & SC 23  & SC 24 & of change \\ %\cline{2-4} 
  & [Number] & [Number] &  $[\%]$ \\\cline{2-4}
   & & &\\ 
 SARs  &11523  & 5607 & \textbf{51} \\
& & &\\ 
 \hline 
 & Peak CARs  & Peak CARs & Rate of change \\ 
 & SC 23  & SC 24 & $[\%]$\\ \cline{2-4} 
 & & &\\
 CARs & 1476 & 1258 & \textbf{15}\\
 & & &\\
 \hline             
\end{tabular}
\end{table}

A comparison of the two panels of Fig. \ref{fig:SARs_CARs} indicates that the monthly number of SARs reaches a maximum during the first peak of the NSN  in both SCs 23 and 24, the same trend as the monthly number of $\beta$ regions (see the upper panel of Fig. \ref{fig:Monthly}). On the other hand, the monthly number of CARs rises to its maximum later, during the second NSN peak in both cycles. A similar pattern was also observed in the monthly numbers of $\beta\gamma$ and $\beta\gamma\delta$ regions (the lower panel of Fig. \ref{fig:Monthly}).

\begin{figure}[h]
\resizebox{\hsize}{!}{\includegraphics{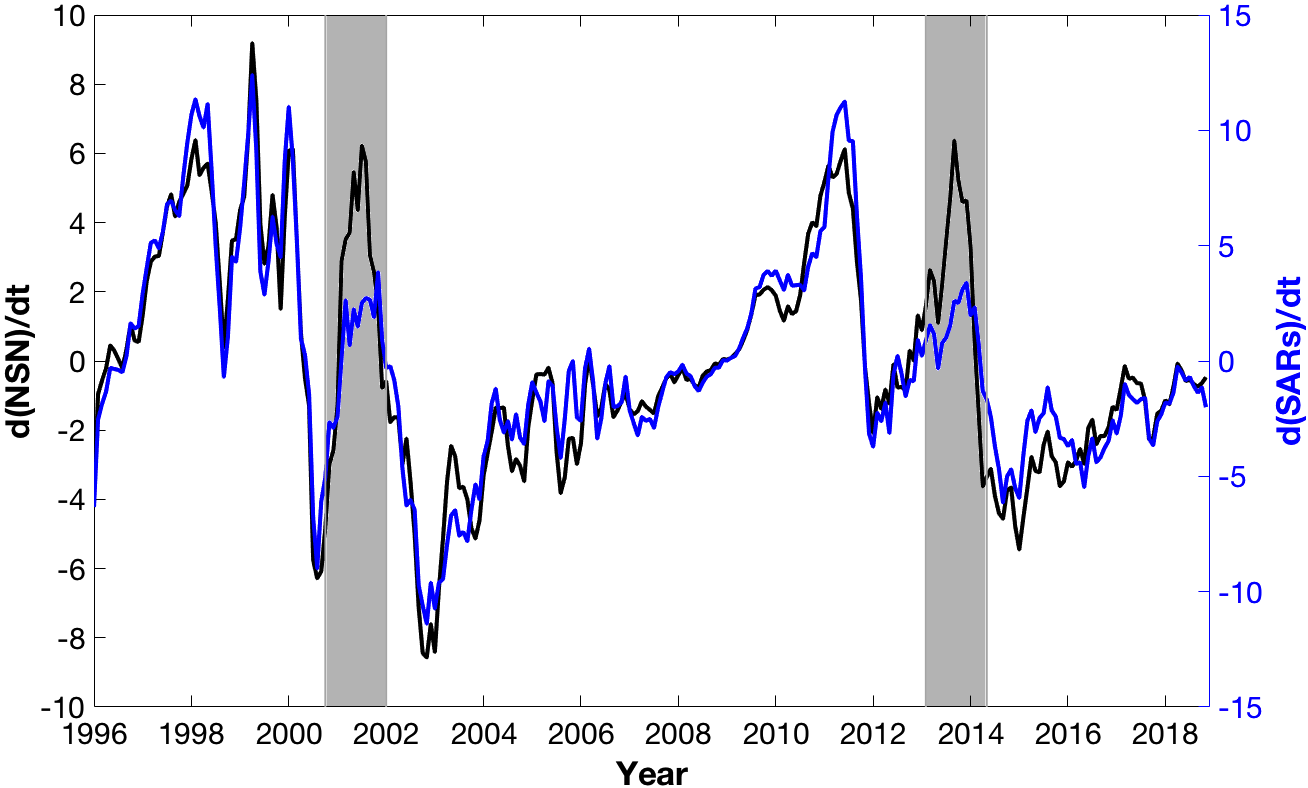}}
\resizebox{\hsize}{!}{\includegraphics{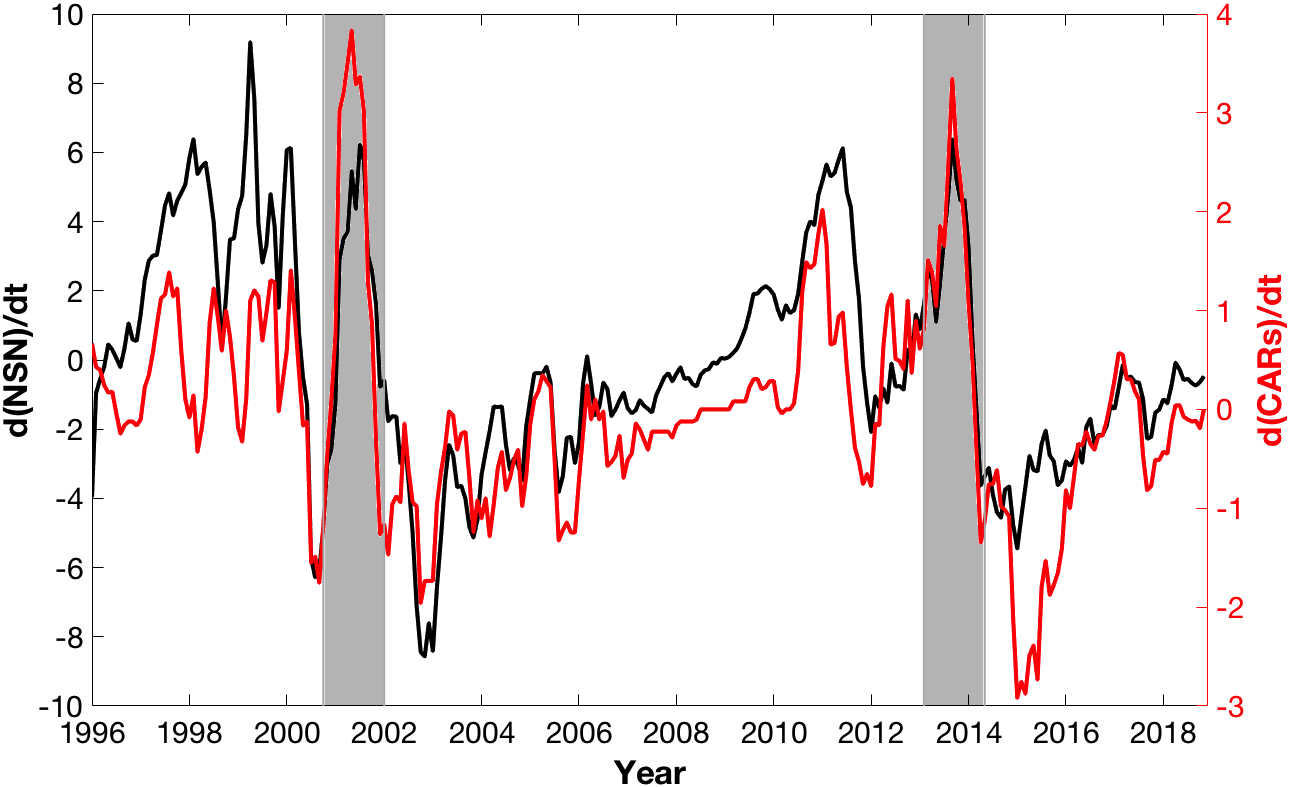}}
\caption{Upper panel: Time variation of SARs (blue line) and the NSN (black line). Lower panel: Time variation of CARs (red line) and the NSN (black line). Gray shaded bars show the periods when the largest difference between the time variation of SARs and the NSN happens.}
\label{fig:Time}
\end{figure}

We also analyzed the monthly number of SARs and CARs for the periods in which they achieved their maximum values in both SCs 23 and 24. In order to do this, we calculated the maximum of SARs and CARs separately for each cycle. The maximum of SARs was reached in May 2000 during cycle 23 and in December 2011 during cycle 24,  while CARs peaked in November 2001 and January 2014 during cycle 23 and 24, respectively (these maximum values are marked with circles in Fig. \ref{fig:SARs_CARs}). Then, we counted the number of SARs and CARs for a period of four years: two years on either side of their maximum values in both cycles. Table \ref {tab:maxima} presents the results from this analysis, together with the rate of change of SARs and CARs from SC 23 to 24 during their peak periods. The data in Table \ref {tab:maxima} indicate that the number of SARs dramatically decreased by $51 \%$  from the SARs peak in SC 23 to the SARs peak in SC 24. The number of CARs dropped by only $15 \%$ from the CARs peak in SC 23 to SC 24.  A similar trend was noted for the full cycle, with drop rates of $45\% $ and $ 32\%$ for SARs and CARs, respectively.

Next, we computed the time derivative of the monthly SARs, CARs, and the NSN from January 1996 to December 2018. The smoothed values (seven-months running average) are presented in Fig. \ref{fig:Time}. The upper panel shows that the rate of change in SARs closely follows the variation of the NSN (r=0.99). In each cycle, the largest difference between the variation in SARs and the NSN  occurs when the variation in CARs reaches its maximum. These periods are shown with shaded bars in Fig. \ref{fig:Time}. Moreover, a comparison of the two panels of Fig. \ref{fig:Time} shows that the highest variation in the monthly number of CARs occurs after the highest variation in SARs in each cycle,  as already shown for the monthly numbers of SARs and CARs (see Fig. \ref{fig:SARs_CARs}). 

\begin{figure}[h]
\centering
\resizebox{\hsize}{!}{\includegraphics{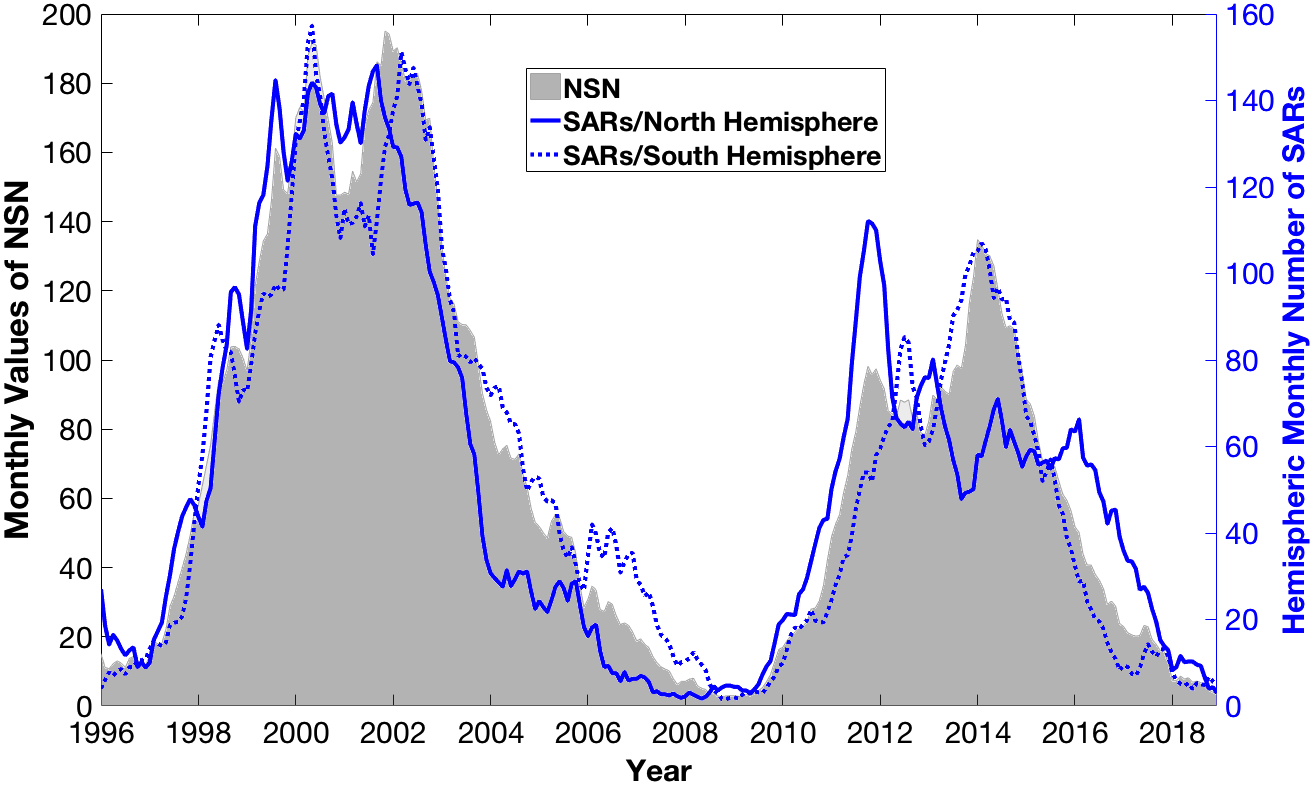}}
\resizebox{\hsize}{!}{\includegraphics{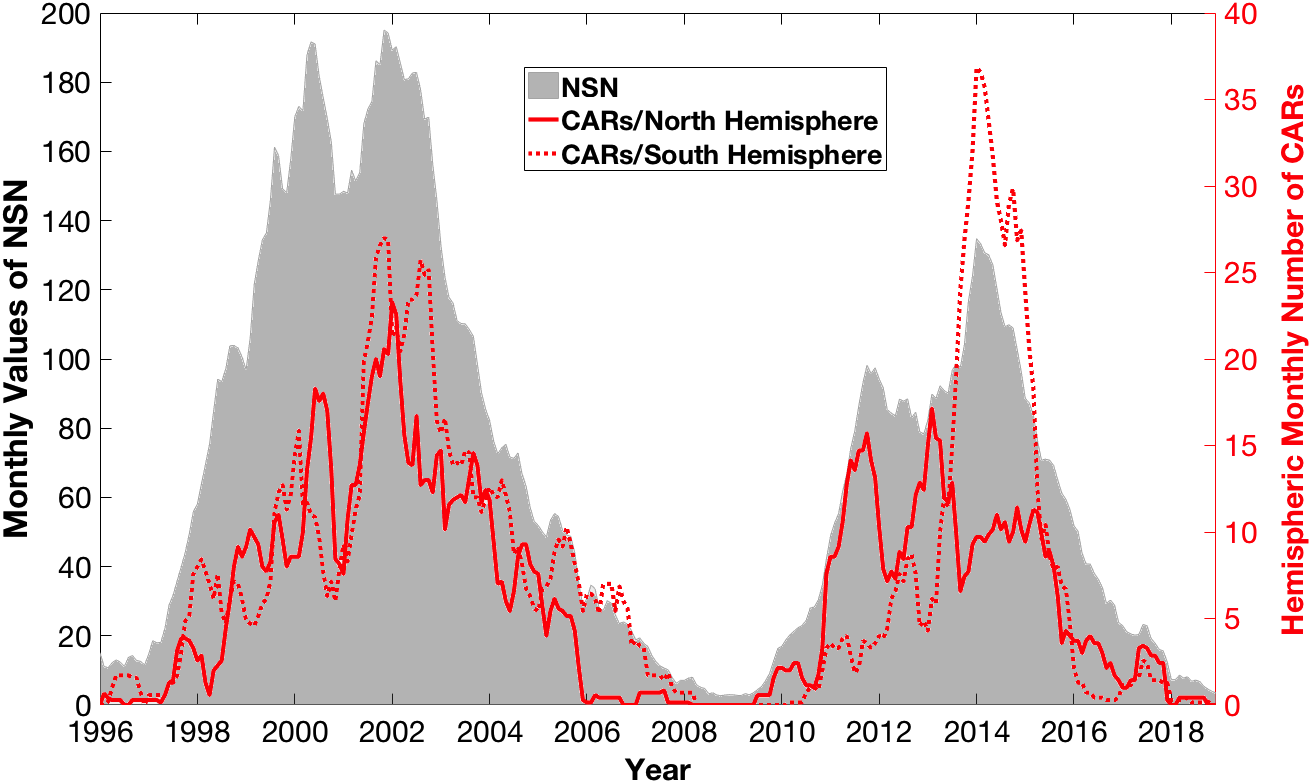}}
\caption{Monthly number of SARs and CARs in the north and south solar hemispheres. The monthly average values of the NSN are illustrated with the gray shaded areas in both panels. Upper panel: Monthly number of SARs in the north and south hemispheres are shown with the solid and dotted lines, respectively. Lower panel: Monthly  number of complex ARs in northern hemisphere (solid line) and southern hemisphere (dotted line).}
\label{fig:Hemispheric}
\end{figure} 

\subsection{North-south asymmetry of active regions}
We calculated the monthly numbers of SARs and CARs in the northern and southern hemispheres separately in order to study the north-south asymmetry of ARs. Figure \ref{fig:Hemispheric} presents the hemispheric monthly number of SARs (upper panel) and CARs (lower panel) together with the monthly value of the NSN. All data have been smoothed using a seven-month moving average). The upper panel of Fig. \ref{fig:Hemispheric} shows a double peak pattern in the monthly number of SARs in the southern hemisphere during the maximum phase of SC 23, whereas the number of SARs is relatively constant in the northern hemisphere during the same phase. The number of SARs shows large hemispheric asymmetry in SC 24: the values in the northern hemisphere reach a maximum during the first peak of the NSN, whereas the values in the southern hemisphere peak later, during the second peak of the NSN. 

In the lower panel of Fig. \ref{fig:Hemispheric}, the monthly number of CARs in the northern and southern hemispheres are presented against the monthly value of the NSN. As can be seen, there is a double peak pattern in the number of CARs in both hemispheres during the maximum phase in SC 23 while the maximum of CARs was achieved during the second peak of the NSN. Similar to SARs, the CARs number shows an asymmetry during the maximum phase in SC 24: the number of CARs in the northern hemisphere reaches a maximum during the first NSN peak, while in the southern hemisphere the maximum is reached during the second peak.

\begin{figure}[h]
\resizebox{\hsize}{!}{\includegraphics{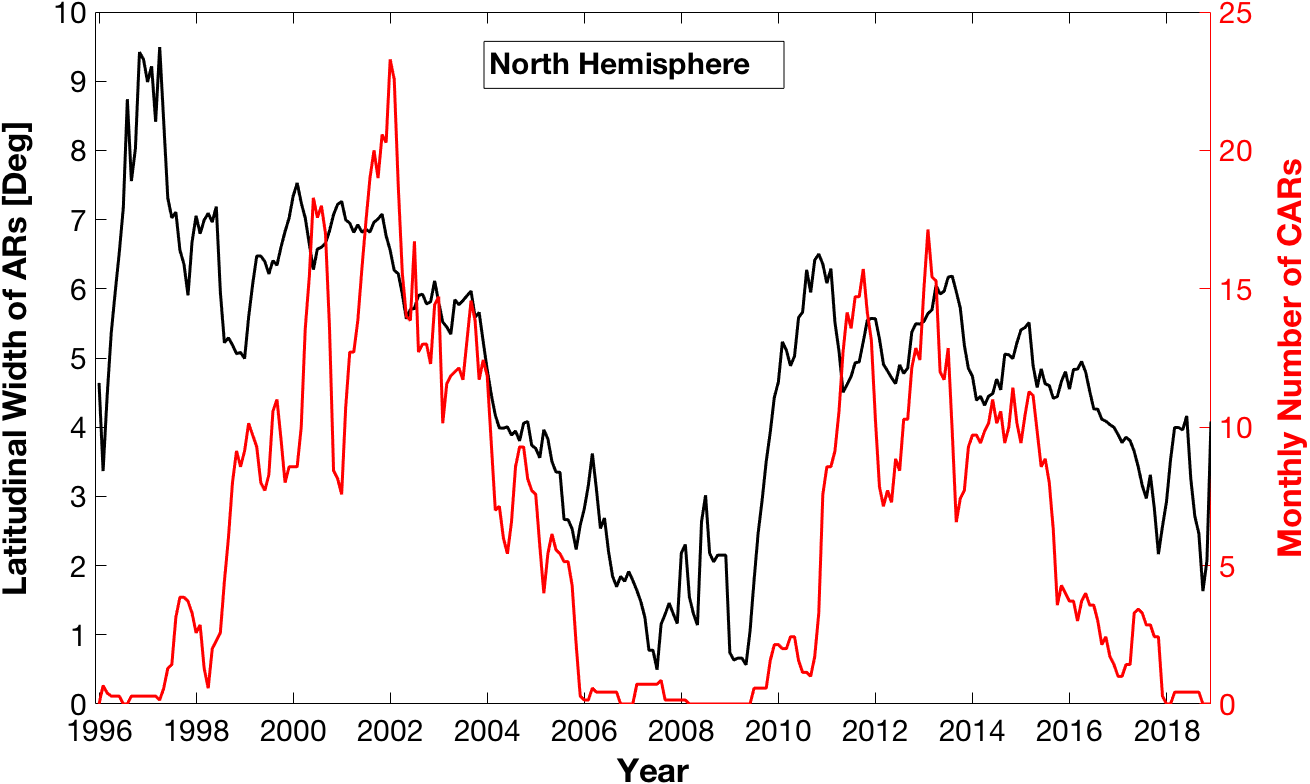}}
\resizebox{\hsize}{!}{\includegraphics{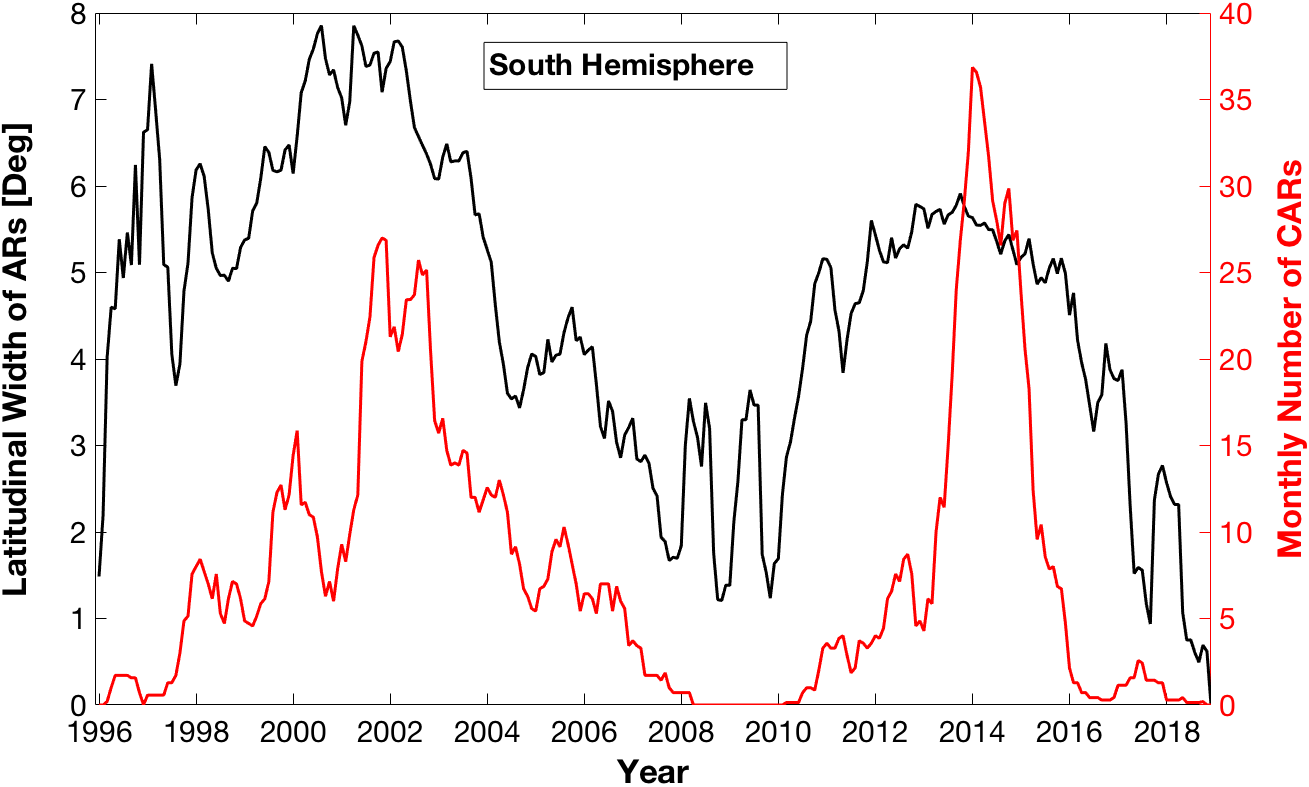}}
\caption{Comparison between the latitudinal width of ARs and the monthly number of CARs in the solar northern hemisphere (upper panel) and southern hemisphere (lower panel). The black and red lines present the latitudinal width of ARs and the monthly number of CARs, respectively.}
\label{fig:Latwidth}
\end{figure}

\subsection{Latitudinal distribution of active regions}
\label{subsec:latitude}
We conducted an analysis on the latitudinal distribution of all ARs in each hemisphere for both SCs 23 and 24. The latitudinal width of ARs was computed by using monthly values in order to compare it with the monthly number of CARs. Figure \ref{fig:Latwidth} presents the results of this comparison for the northern hemisphere (upper panel) and the southern hemisphere (lower panel). The latitudinal width of ARs increases during the ascending phase and it is relatively constant during the maximum phase in both hemispheres during SCs 23 and 24. However, during both cycles, the latitudinal width of ARs remains relatively constant until CARs reach their maximum, and only after that does the width start to decrease.

\begin{figure}[h]
\resizebox{\hsize}{!}{\includegraphics{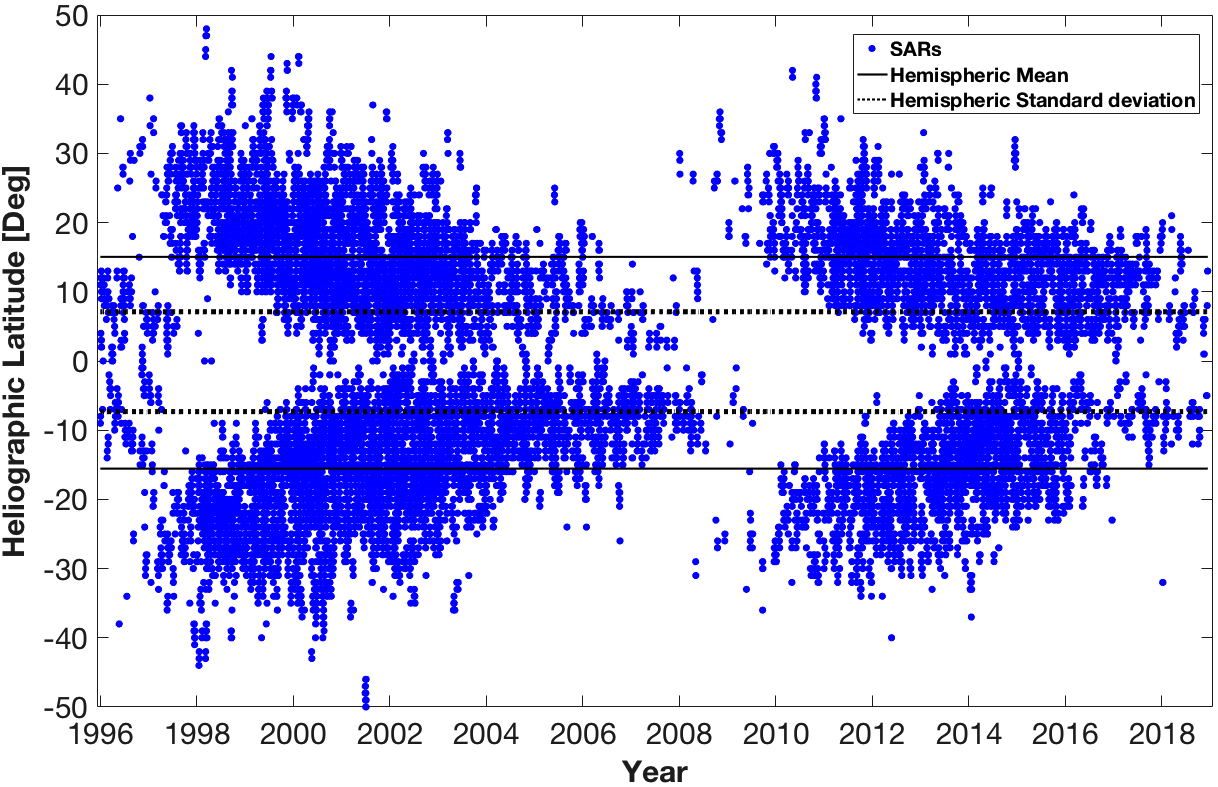}}
\resizebox{\hsize}{!}{\includegraphics{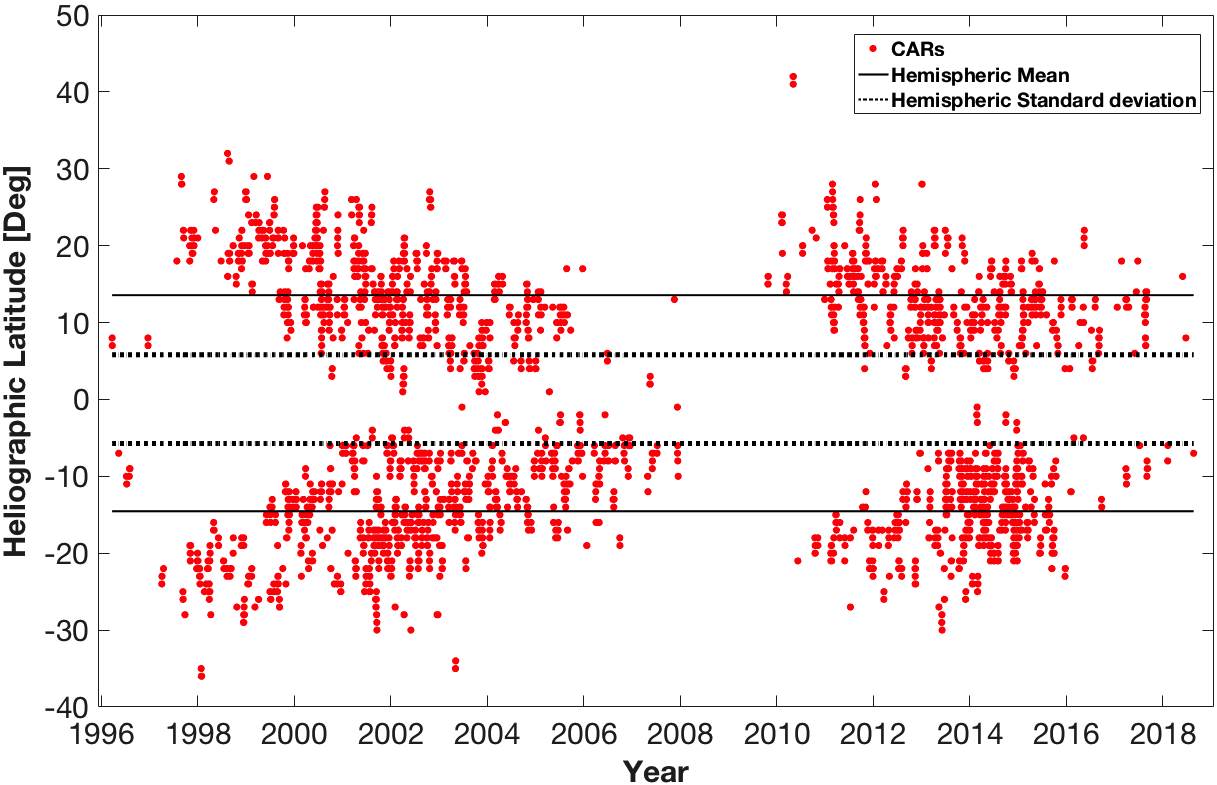}}
\caption{Daily latitudinal variation of SARs (upper panel) and CARs (lower panel) from January 1996 to December 2018. Hemispheric mean and standard variation are shown with black solid and dashed lines, respectively.}
\label{fig:Latitude}
\end{figure}

We also calculated the distribution of heliographic latitudes of SARs and CARs using the daily data for the period of January 1996 to December 2018. The results are presented in Fig. \ref{fig:Latitude} (the solid and dotted lines are the hemispheric mean and  standard deviation, respectively).  Figure \ref{fig:Latitude} shows that in both SCs 23 and 24 almost all CARs emerged between the latitudes $-30^{\circ}$ and $30^{\circ}$, whereas SARs appeared in a wider latitudinal band (or more precisely, between $-40^{\circ}$ and $40^{\circ}$).

\begin{figure*}
\subfloat{\includegraphics[height=0.22\textheight]{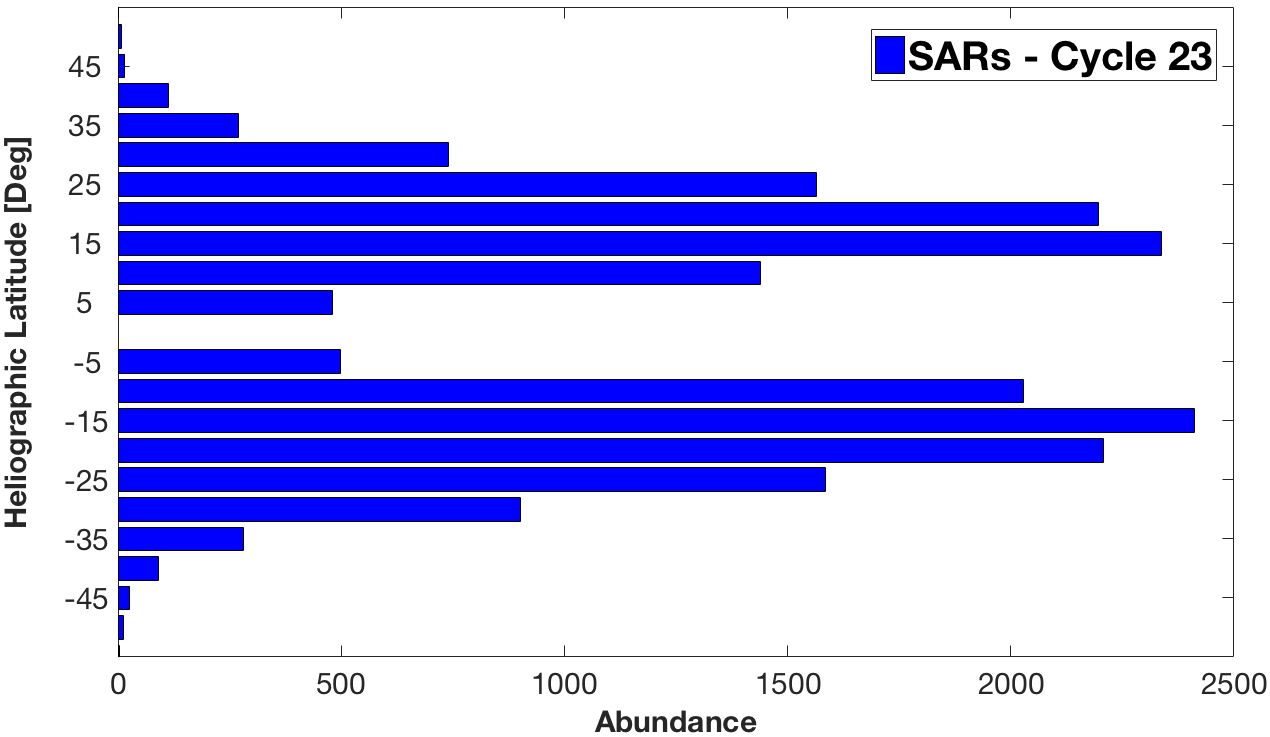}}
\subfloat{\includegraphics[height=0.22\textheight]{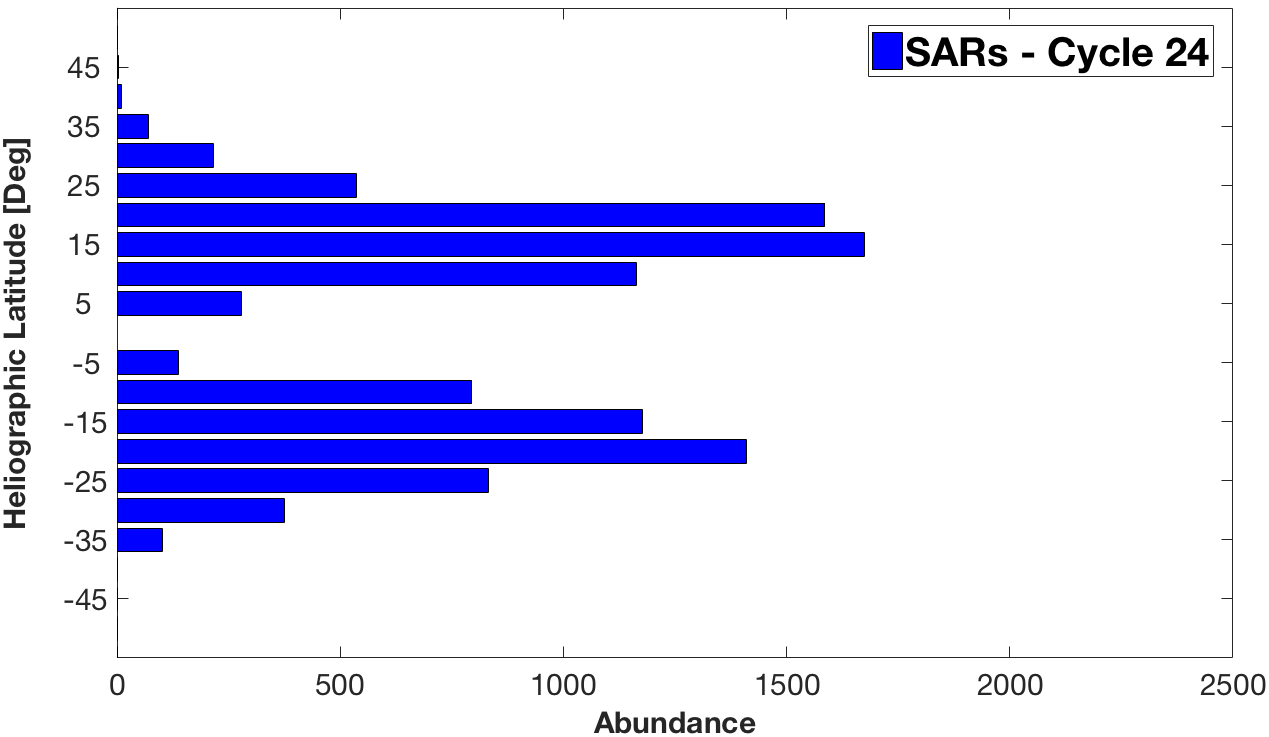}}\\
\subfloat{\includegraphics[height=0.22\textheight]{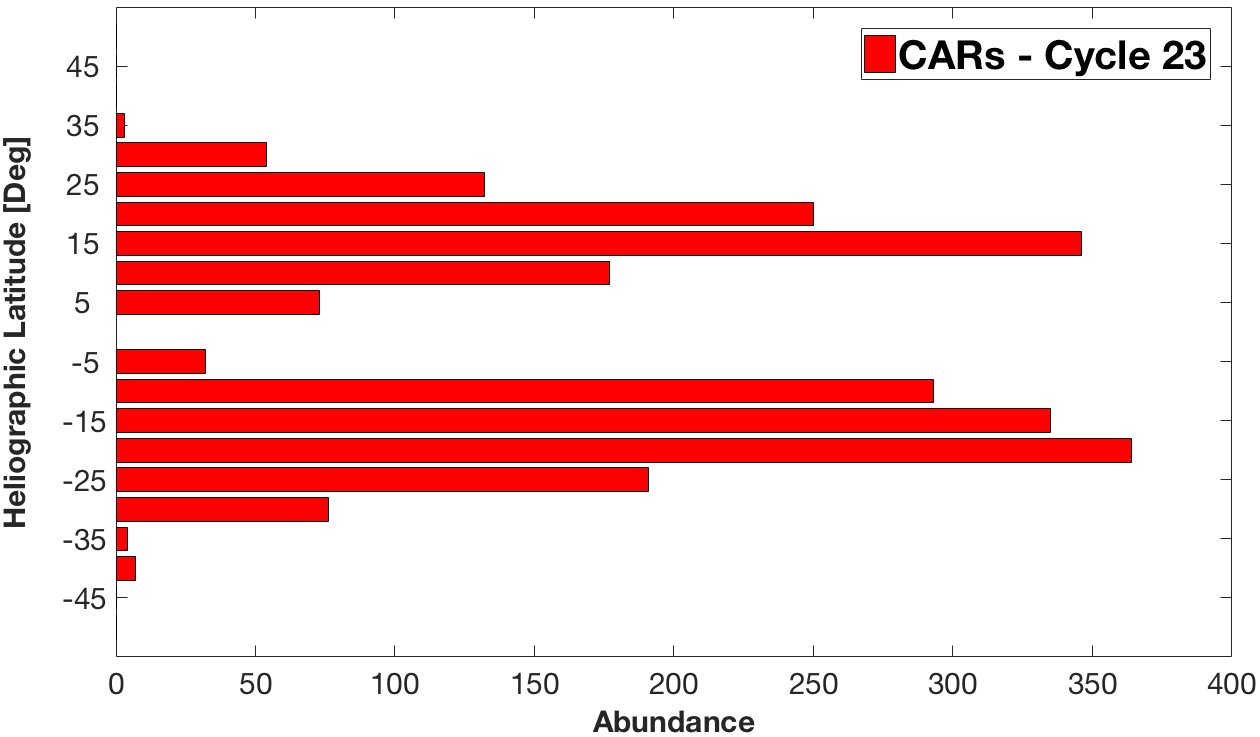}}
\subfloat{\includegraphics[height=0.22\textheight]{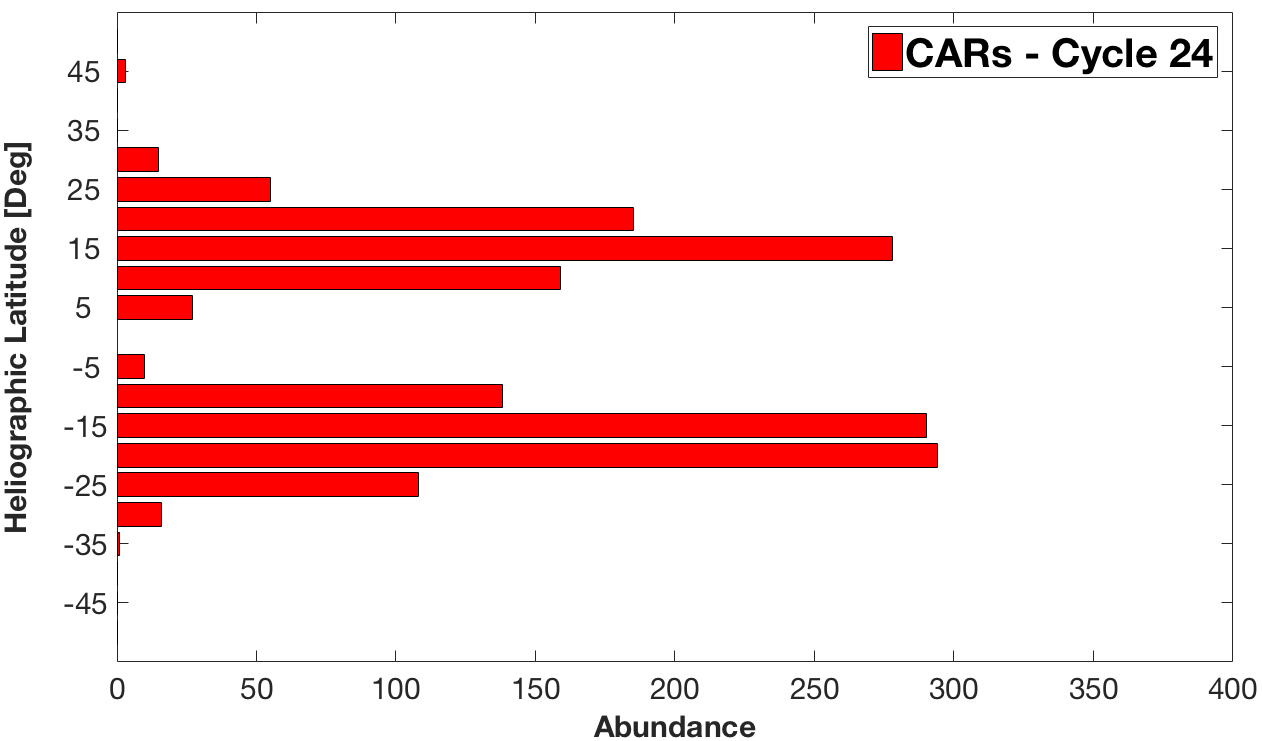}}
\caption{Latitudinal distribution of SARs (blue bars) and CARs (red bars) in the northern and southern hemisphere during SCs 23 and 24. In these plots, each bar represents a latitudinal band of five-degree width.}
\label{fig:Bars}
\end{figure*}

We then performed the two sample Kolmogorov-Smirnov (KS) test \citep{Numerical2007} in order to study the differences between the latitudinal distribution of SARs and CARs in the northern and southern hemispheres during SCs 23 and 24. The KS test for SC 23 showed that the probabilities of the latitudinal distribution of SARs and CARs being part of the same sample were $P = 0.54$ and $P = 0.39$ in the northern and southern hemispheres, respectively. The results from a similar analysis for SC 24 showed that the probabilities that the latitudinal distribution of SARs and CARs belong to the same population in the northern hemisphere is $P = 0.74$ and $P = 0.95$ in the southern hemisphere.

We also investigated the latitudinal distribution of SARs and CARs in the northern and southern hemispheres during SCs 23 and 24. We divided each hemisphere into five-degree latitudinal bands and then computed the number of SARs and CARs within the bands. Figure \ref{fig:Bars} presents the latitudinal distribution of SARs (upper panels) and CARs (lower panels) in SCs 23 and 24, respectively. A comparison of the two upper panels indicates that the latitudinal distribution of SARs in SC 23 is nearly the same as in SC 24. The latitudinal distribution of CARs (lower panels) in SCs 23 and 24 is fairly similar to that of SARs.

\section{Discussion}
We studied the magnetic complexity of ARs with a new approach that allows us to investigate their daily complexity abundance. The results showed that about $88 \%$ of the entire sample have simple magnetic structures (i.e., $\alpha$ and $\beta$). This matches with the finding of \citet{Amiee2016} who also showed that $84 \%$ of all identified ARs in their sample had simple magnetic structures, while \citet{Hale1938} found a somewhat higher abundance ($96\%$). \citet{Hale1938} studied ARs from 1915 to 1924, covering the solar maximum, minimum, and the declining phase of SC 15, which may explain the difference. We also found that regions with $\beta\delta$, $\beta\gamma$, $\beta\delta\gamma$, $\gamma,$ and $\gamma\delta$ configurations comprised about $12 \%$ of all ARs, which is rather similar to the number ($16 \%$) reported by \citet{Amiee2016}. \citet{Hale1938} reported only two types of complex structures, $\beta\gamma$ and $\gamma$, with a total occurrence rate of $4 \%$. Improvement in observational techniques may explain the increased abundance of complex ARs in this study. 

The monthly number of SARs, shown in Fig. \ref{fig:SARs_CARs}, closely follows the monthly NSN with a significant linear correlation (r= $0.98$), while the monthly number of CARs showed slightly weaker correlation (r= $0.87$). This result is in agreement with that obtained by \citet{Amiee2016}, which displayed a variation in the magnetic complexity of ARs as a function of solar cycle. In contrast, \citet{Hale1919} found no correlation between the complexity variation of ARs and changes in the solar cycle. We also found that the monthly number of SARs attained their maximum during the first peak of the NSN, whereas CARs reached their maximum roughly two years later, during the second peak of the NSN (see Fig. \ref{fig:SARs_CARs}).  We interpret this behavior in terms of the competition between the two different solar dynamo processes. The solar small-scale dynamo (SSD) that generates a fluctuating magnetic field is most likely ubiquitously distributed in the solar convection zone \citep{Batchelor1950}. It is expected and supported by observational data, to be independent of the solar cycle \citep{Petrovay1993, Buehler2013}. 
 The rotational influence on turbulent convection, together with solar differential rotation, is thought to give rise to another dynamo instability, the large-scale dynamo (LSD, \citet{Parker1955}, which dominates in the deeper layers of the convection zone. It gives rise to the cyclic, large-scale component of the solar magnetic field. When LSD is growing, it appears to dominate over the SSD in the AR formation process. Hence the abundance of CARs only starts peaking when the LSD contribution starts declining,  and SSD grows in relative strength, influencing more the AR formation process and evolution. The overall AR formation process, however, is linked to the total intensity of the magnetic field, which is cyclically varying, hence both SARs and CARs eventually stop forming during the solar minimum.

Surprisingly, we found that the total abundance of CARs is very similar during a period of four years around their maximum number in SCs 23 and 24 (Table \ref{tab:maxima}). This result supports our interpretation of this process being related to SSD, which should not show cycle to cycle variation. The behavior of SARs shows that the overall ARs formation process is related to the LSD process.

We showed that the yearly and monthly numbers of $\alpha$ ARs were almost constant during solar maximum in the cycles studied. We hypothesize that $\alpha$ regions are births or deaths of other ARs. This finding agrees with \citet{Bumba1964}, who showed that new ARs appear in or immediately adjacent to older ARs. \citet{Smith1968} also reported that all unipolar (or $\alpha$) regions are old remnants of $\beta$ regions.

The hemispheric monthly number of SARs  showed a double peak pattern in the southern hemisphere during SC 23. This contributes to the double peak pattern seen in the monthly NSN during SC 23. In addition, we found that the monthly number of SARs was asymmetric between the northern and southern hemispheres during SC 24. The number of SARs  in the northern hemisphere reached maximum value during the first peak of the NSN, while the number in the southern hemisphere attained the maximum value more than two years later, during the second peak of the NSN. On the other hand, the monthly number of CARs had a double peak pattern in the northern and southern hemispheres during SC 23, with a maximum during the second peak of NSN. For SC 24 both CARs and SARs showed an asymmetric pattern; the CARs number increased in the northern hemisphere during the first NSN peak, while CARs value in the southern hemisphere reached a maximum later and during the second peak of the NSN.

Comparing the monthly latitudinal width of ARs with the monthly number of CARs in SCs 23 and 24 showed that complex regions  peaked before the latitudinal width starts to decrease. This finding is contrary to the hypothesis of \citet{Amiee2016} who suggested that complex ARs should appear during the declining phase due to the decrease of the latitudinal width.

We demonstrated that the latitudinal distributions of SARs and CARs were almost the same during SCs 23 and 24, but SARs emerged in a wider latitudinal band than CARs in both cycles (see Fig. \ref{fig:Bars}). In addition, applying the two sample KS test to the latitudinal distribution of SARs and CARs showed high probability that SARs and CARs in the northern and southern hemispheres belong to the same population during SC23. A similar analysis for SC 24 showed an even higher probability that SARs and CARs belong to the same population in each hemisphere. This finding suggests that SARs and CARs have the same origin within the solar interior, which confirms the results of \citet{Amiee2016}. We suggest that all ARs are produced at the same depth in the solar interior and some of them are transformed into CARs near to the solar surface.

\section{Conclusions} 
The aim of the present study was to investigate the solar cycle variability of simple active regions (SARs) and complex active regions (CARs), their latitudinal distribution, and north-south asymmetry by using the daily approach developed and presented for the first time in this paper.  By studying SARs and CARs separately, we were able to examine their relative roles over solar cycles 23 and 24 (January 1996 - December 2018). We found differences and similarities between SARs and CARs. The main differences were found in the solar cycle evolution, relative abundance of the simple and complex active regions, and cycle-to-cycle variability of their abundance. The similarities were seen in hemispheric behavior and statistical latitudinal distribution.

Similarities between SARs and CARs were seen in their hemispheric behavior such that they show strong hemispheric asymmetry in solar cycle 24 but both are quite symmetrically distributed in solar cycle 23. The latitudinal distributions of SARs and CARs were similar, although SARs emerged in a slightly wider latitudinal band than CARs in both cycles studied. This finding suggests that SARs and CARs originate from the same location of the convection zone, confirming the results of \citet{Amiee2016}.

The first major difference between SARs and CARs was their relative abundance. We found that the $\alpha-$ and $\beta-$ type ARs account for about $31 \%$ and $58 \%$ of all ARs, respectively. ARs with a complex magnetic structure comprise approximately $12 \%$ of the total sample. This result agrees with \citet{Amiee2016}, who found that slightly over $83 \%$ of ARs are simple and $16 \%$ are complex.

 The second difference between SARs and CARs is in their solar cycle evolution. Our study shows that SARs closely follow the NOAA sunspot number (r=0.99), while CARs reach their maximum number roughly two years later. The majority of CARs emerged during and after the second peak of the cycle. Since SARs cover almost $90 \%$ of all ARs, they dominate the solar cycle evolution of ARs. The role of CARs becomes relatively more important when turning to the second half of the solar maximum and early declining solar cycle phase. We tested the \citet{Amiee2016} hypothesis that the number of CARs increases due to the decrease in the latitudinal width. We found that the number of CARs peaked before the latitudinal width started to decrease, and thus the changes in the latitudinal CARs coverage cannot be the primary trigger for the changes in the number of CARs.

The third difference, the cycle-to-cycle variability, was studied by comparing the SAR and CAR abundances during solar cycles 23 and 24. The maximum abundance of CARs was found to be surprisingly similar during both cycles studied. The monthly maximum for both cycles was close to 50 (see Fig. \ref{fig:SARs_CARs}), and the rate of change from cycle 23 to 24 was $15 \%,$ while for SARs it was $51\%$. This finding suggests that the mechanism through which ARs transform to CARs does not show cycle-to-cycle variation. This supports our interpretation that such a mechanism may be related to the small-scale dynamo close to the solar surface.

\section*{Acknowledgments}
We would like to thank the National Oceanic and Atmospheric
Administration (NOAA) and the Heliophysics Integrated Observatory (HELIO) for providing the solar active region and sunspot data. We acknowledge the financial support by the Academy of Finland of the ReSoLVE Center of Excellence (project 307411) and Geoscientific infrastructure G-EPOS (project 293488). This project has also received funding from the European Research Council (ERC) under the European Union's Horizon 2020 research and innovation program (grant agreement n:o 818665). The authors thank the anonymous referee for valuable comments, which helped to improve the study.

\bibliographystyle{aa}
\bibliography{Nikbakhsh}
\end{document}